\documentclass[a4paper,fleqn,usenatbib]{mnras}
\usepackage[T1]{fontenc}
\usepackage{ae,aecompl}

\usepackage{graphicx}	
\usepackage{amssymb}	

\usepackage{url}
\usepackage{color}
\usepackage{journals}

\newcommand {\be}{\begin {equation}}
\newcommand {\ee}{\end {equation}}

\title[Accreting magnetar M82 X-2]
      {Propeller effect in action in the ultraluminous accreting magnetar M82 X-2 }
 \author[Tsygankov et al.]
    {Sergey~S.~Tsygankov,$^{1}$\thanks{E-mail: sergey.tsygankov@utu.fi} 
    Alexander A. Mushtukov,$^{1,2}$ 
     Valery F. Suleimanov,$^{3,4}$  
      \newauthor and Juri~Poutanen$^{1,5}$
  \\
$^1$Tuorla Observatory, Department of Physics and Astronomy, University of Turku, V\"ais\"al\"antie 20, FI-21500 Piikki\"o, Finland\\
$^2$Pulkovo Observatory of the Russian Academy of Sciences, Saint Petersburg 196140, Russia\\
$^3$Institut f\"ur Astronomie und Astrophysik, Universit\"at T\"ubingen, Sand 1, D-72076 T\"ubingen, Germany\\
$^4$Kazan (Volga region) Federal University, Kremlevskaja str., 18, Kazan 420008, Russia \\
$^{5}$Nordita, KTH Royal Institute of Technology and Stockholm University, Roslagstullsbacken 23, SE-10691 Stockholm, Sweden
}

\date{Accepted 2016 January 5.  Received 2015 December 15; in original form 2015 July 29}

\pubyear{2015}

\begin{document}
\label{firstpage}
\pagerange{\pageref{firstpage}--\pageref{lastpage}}
\maketitle

\begin{abstract}
We present here the first convincing observational
  manifestation of a magnetar-like magnetic field in an accreting neutron
  star in binary system -- the first pulsating ultra-luminous X-ray
  source  X-2 in the galaxy M82. 
  Using the {\it Chandra} X-ray observatory data we show that 
  the source exhibit the bimodal distribution of the luminosity with two well-defined peaks 
  separated by a factor of 40. 
  This behaviour can be interpreted as the action of the  ``propeller regime'' of accretion. 
  The onset of the propeller in a 1.37~s pulsar at luminosity of $\sim10^{40}$~erg~s$^{-1}$ 
   implies the dipole component of the neutron star magnetic field of $\sim10^{14}$~G. 
\end{abstract}


\section{Introduction}

The revolutionary discovery of the pulsating ultra-luminous X-ray
source (ULX) X-2 in the galaxy M82 (known also as X42.3+59;
\citealt{2006ApJ...646..174K}) made by the {\it NuSTAR} observatory
\citep{2014Natur.514..202B} brought more questions than answers on the
physics of accretion onto magnetized neutron stars as well as on the
nature of ULXs.  The chief distinction of this source is an extremely
high luminosity for an accreting neutron star of about
$10^{40}$~erg~s$^{-1}$. Other parameters of the system, such as the
spin period of $P=1.37$~s, the orbital period of 2.5~d and the
companion star of 5.2~M$_{\odot}$ \citep{2014Natur.514..202B}, are
quite typical for an accretion powered X-ray pulsar in a high-mass
X-ray binary.  The variable over the {\it NuSTAR} observations spin-up
was also detected with an averaged rate $\dot
P\simeq-2\times10^{-10}$~s~s$^{-1}$.

Exceeding the Eddington luminosity by almost two orders of magnitude
is a challenge for current theories and imposes strong limitations on
the physical conditions in the emitting region. Many different models
are already proposed aiming at the understanding of the nature of this
source. These models can be divided into two main groups depending on
the neutron star magnetic field.  One of the possibilities to emit
such a high flux is to assume a magnetar-like magnetic field of the
order of $10^{14}$ G to significantly reduce the interaction
cross-section between radiation and the infalling material \citep[see,
  e.g.,][]{2006RPPh...69.2631H}.  However, such a magnetic field is
almost two orders of magnitude higher than the typical value observed
in X-ray pulsars \citep{2015A&ARv..23....2W} and cannot be probed
through spectral analysis. Therefore, all conclusions made so far
are done based on the timing properties of M82 X-2. Particularly,
assuming torque equilibrium, \citet{2015MNRAS.448L..40E} got the
$B$-field strength of $(2\div 7)\times
10^{13}$~G. \citet{2015MNRAS.449.2144D} favoured a lower magnetic
field of $B\sim10^{13}$~G, solving the torque equation numerically.

\begin{table*}
\caption[{\it Chandra} observations]{{\it Chandra} observations of ULX M82 X-2.} \label{obs_all} 
\centering
\begin{minipage}{130mm}
\begin{tabular}{lccccccc}
\hline\hline
Obs Id &  Instrument & Date &  Exposure & Pile-up fraction & $\alpha$  & Photon  &  Luminosity$^{a}$  \\
       &             &(MJD) &  (ks)     &  (per cent)      &           & index   & ($10^{38}$ erg s$^{-1}$) \\
\hline
361       & ACIS-I & 51441.474 & 33.25  & 1  & -              & $2.1\pm0.4$  & $2.6\pm0.5$ \\ 
1302      & ACIS-I & 51441.883 & 15.52  & 1  & -              & $1.3\pm0.7$  & $2.0\pm0.6$ \\ 
1411\_000 & HRC    & 51479.184 & 36.04  & -  & -              & $0.4\pm1.2$  & $114.3\pm11.3$ \\ 
1411\_002 & HRC    & 51563.619 & 17.61  & -  & -              & 1.35$^{b}$    & $1.7\pm1.0$ \\ 
2933      & ACIS-S & 52443.784 & 18.03  & 20 & $0.24\pm0.21$  & $1.4\pm0.2$  & $60.9\pm5.4$ \\ 
5644      & ACIS-S & 53599.038 & 68.14  & 5  & -              & $1.4\pm0.1$  & $107.4\pm3.4$ \\ 
6361      & ACIS-S & 53600.664 & 17.45  & 5  & -              & $1.3\pm0.1$  & $105.1\pm3.3$ \\ 
8189      & HRC    & 54109.345 & 61.29  & -  & -              & 1.35$^{b}$    & $3.2\pm1.7$ \\
8505      & HRC    & 54112.093 & 83.22  & -  & -              & 1.35$^{b}$    & $3.4\pm1.5$ \\ 
10542     & ACIS-S & 55006.163 & 118.61 & 1  & -              & $1.0\pm0.2$  & $3.8\pm0.4$ \\
10543     & ACIS-S & 55013.936 & 118.45 & 1  & -              & $0.7\pm0.4$  & $2.3\pm0.4$ \\ 
11104     & ACIS-S & 55364.136 & 9.92   & 30 & $0.39\pm0.09$  & $1.1\pm0.2$  & $119.5\pm9.6$ \\ 
13796     & ACIS-S & 56148.648 & 19.81  & 30 & $0.23\pm0.11$  & $1.1\pm0.1$  & $173.8\pm18.4$ \\ 
15616     & ACIS-S & 56347.964 & 2.04   & 5  & -              & 1.35$^{b}$    & $6.4\pm2.3$ \\ 
16580     & ACIS-S & 56691.841 & 46.85  & 30 & $0.43\pm0.23$  & $1.2\pm0.1$  & $187.4\pm24.4$ \\
\hline
\end{tabular}
\begin{flushleft}{
$^{a}$ Luminosity in the 0.5--10 keV energy range corrected for interstellar absorption and 
assuming distance to the source $D=3.3$ Mpc \citep{2014MNRAS.443.2887F}.\\
$^{b}$ Fixed at the averaged value obtained in the bright observations least affected by the pile-up effect (Obs ID 5644 and 6361).
  }\end{flushleft} 
\end{minipage}
\end{table*}

At the same time other authors declare a low magnetic field in M82 X-2
based on the same timing properties of the source. For instance,
\citet{2015MNRAS.448L..43K} argue that the observed torque is
consistent with the accretion disc extending down to the vicinity of
the neutron star surface. This can be the case only if the dipole
magnetic field of the star is low, $B\lesssim10^9$~G. In the original
paper by \cite{2014Natur.514..202B}, the magnetic field of the order
of $10^{12}$~G was estimated under the assumptions of the spin
equilibrium and the Eddington accretion rate.

Thus, 
the lack of an obvious observational manifestation of the strong magnetic
field precludes from any final conclusion.  Here, using the {\it
  Chandra} observatory data, we show that the ULX M82 X-2 regularly
enters the ``propeller regime'' of accretion which is seen as dramatic
variations of the emitted luminosity.  These observations imply the
dipole component of the neutron star magnetic field of $\sim10^{14}$~G, 
independently confirming the  magnetar nature of this ULX.

\section{{\it Chandra} observations and results}

The {\it Chandra} X-ray observatory monitored the galaxy M82 more or
less evenly during the past $\sim15$ years resulting in 29 publicly
available observations. Some of them were pointed quite far from the
ultra-luminous X-ray source M82 X-2.  For our analysis, we selected
only on-axis observations where the PSF shape allowed us to
confidently separate the flux from M82 X-2 from the nearby
sources. To exclude the selection bias we checked that none of
  the observations was triggered on a particular state of this
  source. The final sample of the utilized observations consisting of
15 pointings is listed in Table~\ref{obs_all}, where observation Id,
name of the instrument, date and exposure of observation, as well as
pile-up fraction are given. The level of pile-up was estimated
  using the CIAO tool PILEUP\_MAP, which calculates an average number
  of counts per ACIS frame. These numbers were then used to estimate
  the pile-up
  fraction.\footnote{\url{http://cxc.harvard.edu/ciao/download/doc/pileup\_abc.pdf}}

The reduction of the data has been done following the standard
pipeline in the Chandra Interactive Analysis of Observations software
package (CIAO, version 4.7).  The source flux was estimated from the
energy spectrum extracted using CIAO SPECEXTRACT tool from a circular
region of radius $\sim$1\arcsec. The background was extracted
  from an annulus region centered at the source position with the inner and
  outer radii 1\farcs8 and 7\farcs0, correspondingly. All contaminating
  point sources were excluded.

Spectral fitting was performed with XSPEC v. 12.8.1g \citep{Arn96}
using a simple power-law model with interstellar absorption ({\sc
  phabs} model). Spectra were groupped to have a minimum of 1 count
per bin and the Cash statistic was used. Some observations in the
high-luminosity state suffer from the effect of pile-up with a pile-up
fraction around 30 per cent at worse. This makes a reliable spectral
fitting difficult, also taking into account the photon index and
absorption values degeneracy in the {\it Chandra} energy band. The
same is valid for low-luminosity states where we do not have enough
counts to constrain spectral parameters independently. To solve this
problem, we have selected observations in bright states which are much
less affected by pile-up (Obs ID 5644 and 6361) and obtained the value
of the hydrogen column density, $N_{\rm H}=3.11\times10^{22}$
cm$^{-2}$, and used this value to fit all remaining spectra allowing
only the photon index and normalization to vary. Such an approach is
supported by the recent work by \citet{2015arXiv150706014B} where a
detailed spectral study of M82 X-2 was performed and no significant
variability of $N_{\rm H}$ and $\Gamma$ values between different {\it
  Chandra} observations was found. Additionally we checked the
consistency of the spectral shape in different intensity states by the
simultaneous fitting of all spectra obtained in the low state. The
resulting spectral parameters are in very good agreement with ones in
the bright state.  For fitting the piled up spectra from
  observations with the pile-up fraction higher than 5\% we followed
the {\it Chandra} ABC guide to pile-up,$^{\textcolor{blue}1}$
namelly we added the {\sc pileup} model \citep[see
  also][]{2015arXiv150706014B}. The best-fit $\alpha$ parameter,
characterizing the grade morphing, as well as the photon index are
shown in the Table~\ref{obs_all}.

\begin{figure*}
\centering 
\includegraphics[width=12cm]{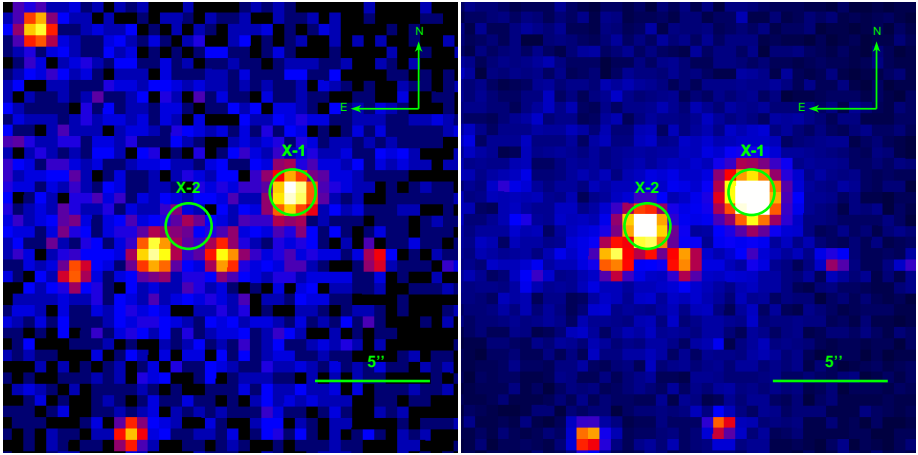} 
\caption{ {\it Chandra} images of M82 galaxy's centre during observations
  performed on September 20, 1999 (MJD 51441.47) when M82 X-2 was in a
  low-luminosity state (left) and August 17, 2005 (MJD 53599.04) when
  it was in a high-luminosity state (right). Circles indicate the
  positions of M82 X-1 and X-2 ultra-luminous X-ray sources. }\label{fig:1}
\end{figure*}

The flux value and its $1\sigma$ uncertainty were calculated using
{\sc cflux} model from the XSPEC package with the fixed hydrogen
column density value.  The resulting luminosity in the energy range
0.5--10 keV corrected for the absorption is shown in the
Table~\ref{obs_all}.  The luminosities of M82 X-2 measured by us are
in a reasonable agreement with the recent results by
\citet{2015arXiv150706014B}, especially if one takes into account the
density of point sources and the presence of the diffuse emission in
the region around X-2 (see Fig.~\ref{fig:1}).

In order to estimate the bolometric flux, we 
assumed the spectral shape of M82 X-2 to be typical to that of X-ray pulsars (power
law modified by a high-energy cutoff at $\sim15$ keV with folding
energy $\sim15$ keV; see, e.g., \citealt{2005AstL...31..729F}), 
resulting in a bolometric correction factor of 2.   
This factor is consistent with the broadband spectrum of the pulsed emission from M82 X-2 as seen by
{\it NuSTAR} \citep{2015arXiv150706014B}.


\begin{figure*}
\centering 
\includegraphics[width=16cm,bb=20 450 570 680, clip]{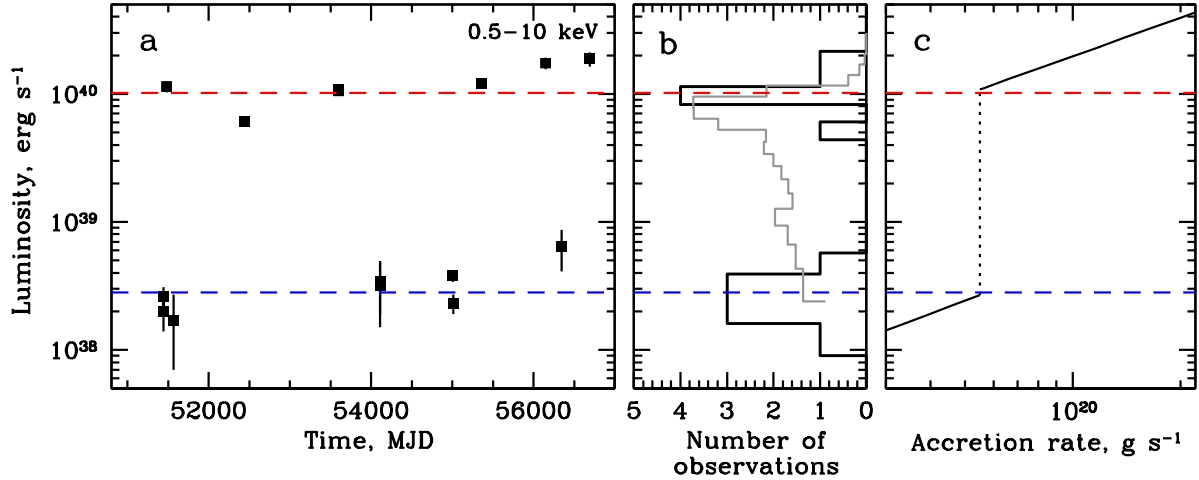} 
\caption{ (a) Light curve of M82 X-2 obtained by the {\it
    Chandra} observatory during 15 years of observations.  Luminosities
  are corrected for the absorption and given in the energy range
  0.5--10 keV;  (b) Distribution of individual
  observations over luminosities (black line). Bimodal structure is
  clearly seen. Red and blue dashed lines show the averaged
  luminosities in the ``high'' and ``low'' states, respectively. Grey
  line represents rescaled distribution of luminosities of X-ray pulsar LMC X-4
  from the {\it Swift}/BAT data; (c) Predicted dependence
  of the magnetised neutron star luminosity on the mass accretion rate
  for the following parameters: spin period $P=1.37$~s, magnetic field
  strength $B=1.1\times10^{14}$~G, neutron star mass
  $M=1.4$M$_{\odot}$ and radius $R=10$~km, magnetospheric radius in
  units of the Alfv\'enic one $k=0.5$.  A small fraction (2.5 per cent) of
  the accreting material is assumed to leak through the magnetosphere
  onto the neutron star surface. 
   }\label{fig:2}
\end{figure*}

The light curve of M82 X-2 as observed by {\it Chandra} is shown  in Fig.~\ref{fig:2}(a). 
The histogram of the luminosities shown in Fig.~\ref{fig:2}(b) clearly demonstrates a bimodality, 
with two well defined peaks at $\sim1.0\times10^{40}$~erg~s$^{-1}$ and
$\sim2.8\times10^{38}$~ergs$^{-1}$.\footnote{The over-all bimodal flux distribution 
is confirmed by Fig.~3 from \citet{2015arXiv150706014B}. 
The only flux measurement there fallen between the two states is the {\it Chandra} observation
10545 (MJD 55405), where the PSF of M82 X-2 is clearly blended with a nearby source,
and hence is missing in our list of observations.}
We stress here, that in the majority of low-luminosity states
the source is still presented in the {\it Chandra} images. This can be
illustrated by Fig.~\ref{fig:1}, where the maps of the central part of
the galaxy M82 are shown in both ``high'' and ``low'' states.

\section{Discussion} 
\subsection{``Propeller'' effect}

The remarkable behaviour of M82 X-2 showing dramatic switches in 
luminosity by a factor of 40 can be interpreted as the onset of
the so-called ``propeller effect'' \citep{1975A&A....39..185I}.  This
effect is caused by a substantial centrifugal barrier which have to be
broken by the infalling matter during the accretion onto the rotating
neutron star with strong magnetic field.  At the magnetospheric radius
where the magnetic pressure equals the ram pressure of the infalling
material, the accreting matter from a disc or a wind is ``frozen''
into the stellar magnetic field lines and rotates rigidly with the
angular velocity of the star.  The matter will fall onto the neutron
star only if the velocity of the magnetic field lines at the
magnetospheric radius is lower than the local Keplerian
velocity. Otherwise, the matter will be stopped at the radius of
magnetosphere or even expelled from the system. Given the fact that
magnetospheric radius depends only on the mass accretion rate and the
strength of the magnetic field, the latter can be directly estimated
if the propeller regime is observed in an accreting magnetized neutron
star.

The threshold value of accretion luminosity $L_{\rm lim}(R)$ for the onset of the
propeller can be estimated by equating the co-rotation radius
(where a Keplerian orbit co-rotates with the neutron star) 
\be
R_{\rm c} = \left( \frac{GMP^2}{4\pi^2} \right) ^{1/3}
\ee
to the magnetospheric radius 
\be
R_{\rm m} 
=k \dot{M}^{-2/7} \mu^{4/7}  (2GM)^{-1/7} . 
\ee
Here $M$ is the neutron star mass,  $P$ its rotational period, 
$\mu=BR^3/2$ is the magnetic dipole moment related to the surface polar dipole magnetic field strength $B$ and 
the neutron star radius $R$, 
and $\dot{M}$ is the mass accretion rate onto the neutron star. 
The dimensionless factor $k$ relates the magnetospheric radius to the 
Alfv\'en radius computed for spherical accretion; 
for  disc accretion it is usually taken $k=0.5$ \citep{GL1978}. 
At the limiting accretion rate $\dot{M}=\dot{M}_{\rm lim}$, $R_{\rm c}=R_{\rm m}$, 
and the accretion luminosity is \citep{2002ApJ...580..389C}   
\be\label{eq1}
L_{\rm lim}(R) \simeq \frac{GM\dot{M}_{\rm lim}}{R} 
\simeq 4 \times 10^{37} k^{7/2} B_{12}^2 P^{-7/3} M_{1.4}^{-2/3} R_6^5 \,\textrm{erg s$^{-1}$} , 
\ee 
where $M_{1.4}$ is the neutron star mass in units of 1.4M$_\odot$, 
$R_6$ is neutron star radius in units of $10^6$~cm,
$B_{12}$ is the magnetic field strength in units of $10^{12}$~G. 

The decrease of the accretion rate below $\dot{M}_{\rm lim}$ 
will lead to the propeller regime of accretion. The accretion efficiency drops significantly and 
the luminosity in that regime corresponds to the accretion onto the magnetosphere with
the maximum value of \citep{1996ApJ...457L..31C} 
\be\label{eq2}
L_{\rm lim}(R_{\rm c}) = \frac{GM\dot{M}_{\rm lim}}{R_{\rm c}} 
= L_{\rm lim}(R) \frac{R}{R_{\rm c}} . 
\ee 
Thus if the pulsar is close to the spin-equilibrium, when $R_{\rm c}\approx R_{\rm m}$, 
small variations in the accretion rate will lead to large variations in the observed luminosity: 
\be\label{eq3}
\frac{L_{\rm lim}(R)}{L_{\rm lim}(R_{c})} = \left(\frac{GMP^2}{4\pi^2R^3}\right)^{1/3} 
\simeq 170 P^{2/3} M_{1.4}^{1/3} R_6^{-1} .
\ee

\begin{figure*}
\centering 
\includegraphics[width=10cm,bb=60 275 565 680]{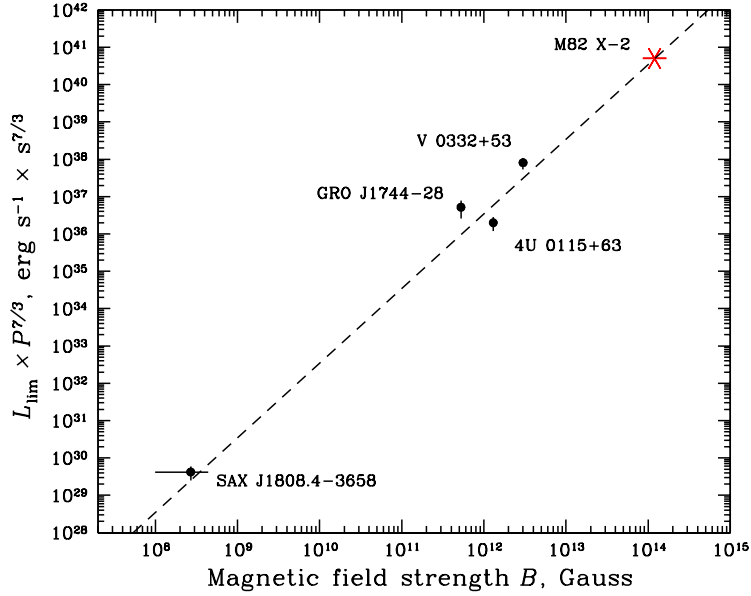} 
\caption{ The observed correlation between the magnetic field strength
  $B$ and a combination of the limiting luminosity at the start of the
  propeller regime and the period, $L_{\rm lim}P^{7/3}$, for four
  pulsars is shown by circles with error bars.  The dashed line
  represents the theoretical dependence from equation~(\ref{eq1}) assuming
  standard parameters $M=1.4 M_{\odot}$, $R=10$~km, $k=0.5$.  The star
  indicates the position of the ULX M82 X-2 for which we estimate
  $L_{\rm lim}P^{7/3}= 5\times 10^{40}$~erg~s$^{4/3}$, that
  corresponds to the $B$-field strength of $1.2\times10^{14}$~G.
}\label{fig:3}
\end{figure*}

Exactly such behaviour, i.e. abrupt switches between two intensity
states, is observed in M82 X-2 (see Fig.~\ref{fig:2}(b)).  To our
knowledge none of the known other variability mechanisms would result
in such a sharp bimodal distribution of flux from an accreting
magnetized neutron star.  For a comparison, we show in
Fig.~\ref{fig:2}(b) by grey line the luminosity distribution for the
X-ray pulsar LMC~X-4 demonstrating a well established super-orbital
variability as seen by the {\it Swift}/BAT all-sky monitor.\footnote{See
  \url{http://swift.gsfc.nasa.gov/results/transients/LMCX-4/}. 
    Daily averaged count rates were translated into the source
    luminosity using its broadband spectrum \citep[see,
      e.g.,][]{2005AstL...31..380T}. The obtained luminosity
    distribution was shifted by  a factor of 20
    to visually match the M82 X-2 luminosity range.}  The variability
patterns obviously differ much. This is also true for other sources
with periodic (SMC~X-1, Her~X-1) and non-periodic (e.g., Cen~X-3)
variability of the flux. Unfortunately, the relatively short exposures
of the {\it Chandra} observations do not permit us to detect
transitions from one state to another.  However, a quite dense
observational program of M82 allows us to put an upper limit on the
duration of such transitions: two observations separated by $\sim37.3$
days (ObsID 1302 and 1411\_000) show flux change by a factor of
$\sim50$.

\subsection{Magnetic field measurement} 

Theoretically, expressions (\ref{eq1})--(\ref{eq3}) would allow us to
measure the magnetic field of the neutron star together with its
compactness.  However, the source does not follow the theory so
closely.  Indeed, the observed ratio of luminosities in ``high'' and
``low'' states is 40, that is smaller than the ratio around 210
  predicted by equation (\ref{eq3}) for a given pulse period,
probably because of a substantial leakage of accreting material
through the magnetosphere.  Such a behaviour is typical for accreting
X-ray pulsars \citep{2011A&A...529A..52D,2014A&A...561A..96D}.  This
is also supported by the fact that we do not detect significant
changes of the source spectra in the two states.  The jump by a factor
of 40 in the luminosity can be easily explained if 2--3 per cent of
accreting matter leaks through the magnetosphere (see
Fig.~\ref{fig:2}c).

However, the limiting luminosity in the high state is less affected by
such uncertainties in the accretion model and can be used to estimate
the magnetic field strength of the neutron star directly from equation
(\ref{eq1}). For this purpose we take $P=1.37$~s and $L_{\rm
  lim}=2.0\times10^{40}$~erg~s$^{-1}$ (where we apply a bolometric
correction of factor 2 to the observed mean luminosity in the high
state).  This gives immediately a magnetar-like magnetic field of
$B=(1.1\pm0.4)\times 10^{14}$~G, which exceeds by a factor of 15 the
largest measured so far magnetic field in an accreting X-ray pulsar
\citep{2013ATel.4759....1Y}.  The error on $B$ comes from the
uncertainty in a combination of factors $k^{7/2}M_{1.4}^{-2/3}R_6^5$
in relation (\ref{eq1}) as well as from a 50 per cent uncertainty in
determination of the limiting luminosity (as a consequence of a
  finite width of the luminosity distribution in ``high'' state) and the
  bolometric correction. For example, if instead of using an averaged
  luminosity in high state as the limiting one we  take the
  minimal observed in this state ($1.2\times10^{40}$~erg~s$^{-1}$), the
  magnetic field strength would be $B=8.5\times 10^{13}$~G.
 
The reliability of this method is illustrated by Fig.~\ref{fig:3},
where we show the correlation between the combination $L_{\rm
  lim}P^{7/3}$ and the magnetic field strength for one accreting
millisecond pulsar SAX J1808.4--3658 and three X-ray pulsars GRO
J1744--28, 4U 0115+63, and V 0332+53 where the action of the propeller
was mentioned in the literature (see Appendix \ref{app:prop}). This
sample consists only of sources with the confidently determined nature
of the compact object as a neutron star with a sufficiently strong
magnetic field, i.e. sources with the pronounced X-ray pulsations.
We see that the data spread over four orders of magnitude
in $B$ are well described by a theoretical dependence given by
equation (\ref{eq1}) and shown by the dashed line in Fig.~\ref{fig:3}.

\section{Conclusion}

In spite of a number of suggestions for the presence of magnetars in
binary systems
\citep{1975SvAL....1..223S,2008ApJ...683.1031B,2010A&A...515A..10D,2012MNRAS.425..595R,2014MNRAS.437.3863K},
no such systems up to date have been unambiguously identified.  Here
we show that the ULX M82 X-2 regularly enters the ``propeller regime''
of accretion which is seen as dramatic variations of the emitted
luminosity and two well-defined peaks separated by a factor of 40 in
the X-ray luminosity distribution.  These observations imply the
dipole component of the neutron star magnetic field of $\sim10^{14}$
G, making the source the first confirmed accreting magnetar.

Our discovery of such a strong magnetic field in a pulsar-ULX naturally
explains the observed deficit of ULXs powered by accretion onto a
neutron star. For instance, binary population synthesis models
suggest \citep{2015ApJ...802..131S} neutron star--ULX to be not less
numerous than black hole--ULX.  Moreover, the observed binary
parameters \citep{2014Natur.514..202B} of M82 X-2 (donor mass $M_{\rm
  c}>5.2M_{\odot}$ and orbital period $\sim2.5$~d) are shown to be
typical.  However, these studies do not account for an extreme
narrowness of all physical parameters permitting for such a binary to
operate as ULX \citep{2015MNRAS.454.2539M}. Namely, the magnetic field
strength should be high enough to make the scattering cross-section
small in order to support super-Eddington luminosity from the
accretion column.  Furthermore, for a high accretion rate needed for a
neutron star to become ULX, the magnetic field should be high enough
to make magnetosphere larger than the spherization radius (where the
accretion disc scale-height becomes comparable to the distance from
the neutron star), otherwise the accreting material will be blown away
by radiation and the accretion may proceed only in a nearly
spherically-symmetric fashion, which is limited by the Eddington
luminosity \citep{2015MNRAS.454.2539M}. This naturally limits the
magnetic field of pulsar-ULXs to be greater than
$\sim3\times10^{13}$~G.

\section*{Acknowledgements}

This work was supported by the Russian Science Foundation grant
14-12-01287 (SST), the Magnus Ehrnrooth Foundation (AAM), the Russian
Foundation for Basic Research 12-02-97006-r-povolzhe-a and the
Deutsche Forschungsgemeinschaft (DFG) grant WE 1312/48-1 (VFS), and
the Academy of Finland grant 268740 (JP).  We also acknowledge the
support from the COST Action MP1304.  The research used the data
obtained from the HEASARC Online Service provided by the NASA/GSFC.

\bibliographystyle{mnras}
\bibliography{allbib}

\appendix

\section{Observational evidence of the onset of the propeller regime}
\label{app:prop}

For Fig.~\ref{fig:3} we have collected all the cases of X-ray pulsars
where the action of the propeller was mentioned in the literature.  We
selected only sources with the confidently determined nature of the
compact object as a neutron star with a sufficiently strong magnetic
field, i.e. sources with the pronounced X-ray pulsations.

The first discovered accreting millisecond pulsar SAX J1808.4--3658
has the shortest spin period of $P=2.5$~ms and the lowest magnetic
field strength among sources in the sample determined to be
$B=(0.8\pm0.5) \times 10^8 k^{-7/4}$~G \citep{2009MNRAS.400..492I}.
Substituting $k= 0.5$ (as we do in our work) the final magnetic field
strength on the neutron star shown in Fig.~\ref{fig:3} is
$B=(2.7\pm1.7)\times10^8$~G.
Variations in the flux by a factor of 100 have been observed by the
{\it Swift} observatory during the 2005 outburst and interpreted as the
transition to the propeller regime of
accretion \citep{2008ApJ...684L..99C}. The luminosity at the onset of
propeller in the 0.3--10 keV energy range was estimated to be
$L=(3\pm1)\times10^{35}$erg~s$^{-1}$ assuming a source distance of
3.5~kpc \citep{2006ApJ...652..559G}. Applying a bolometric correction of
factor 1.7 we get $L_{\rm lim}=(5\pm2)\times10^{35}$~erg~s$^{-1}$.

An intermediate X-ray pulsar GRO J1744--28 situated 1$^{\rm o}$ from
the Galactic centre has the pulse period of 0.467~s and the magnetic
field $B\approx 5.3\times 10^{11}$~G determined from the cyclotron
absorption line at $E_{\rm cyc}\approx
4.7$~keV \citep{2015MNRAS.449.4288D,2015MNRAS.452.2490D}. Because this feature originates
from the vicinity of the neutron star surface, its energy was corrected for the
gravitational redshift to be compared to the magnetic field of
SAX~J1808.4--3658.  The disappearing of strong pulsations from the
source during the decay of 1996 outburst when the 2--60 keV flux was
in the range $F=(4\pm2)\times 10^{-9}$~erg~cm$^{-2}$~s$^{-1}$ was
interpreted as the onset of the propeller
regime \citep{1997ApJ...482L.163C}. Assuming a distance to the source of
8~kpc \citep{1996ApJ...469L..25G}, the propeller threshold luminosity
can be estimated as $L _{\rm lim}=(3.0\pm1.5)\times10^{37}$ erg
s$^{-1}$.

The propeller effect was observed also in two well-studied transient
X-ray pulsars with Be companions -- 4U\,0115+63 and V\,0332+53,
characterized by the spin periods of 3.6~s and 4.35~s and the magnetic
fields of $B\approx 1.3\times10^{12}$~G \citep{1983ApJ...270..711W}
and $3.0\times10^{12}$~G \citep{1990ApJ...365L..59M},
respectively. The sudden decrease of the flux of 4U\,0115+63 and its
pulsed fraction below the luminosity $L_{\rm
  lim}=(1\pm0.5)\times10^{35}$ erg s$^{-1}$ can be interpreted as the
onset of the propeller \citep{2001ApJ...561..924C}.  In the case of
V\,0332+53, the minimum flux in the 1--15 keV range measured just
before the turnoff was
$F=(3\pm1)\times10^{-10}$~erg~cm$^{-2}$~s$^{-1}$
\citep{1986ApJ...308..669S}.  Taking the distance to the source of
7~kpc \citep{1999MNRAS.307..695N} and applying the bolometric correction
of factor 1.5 we get $L_{\rm lim}=(2.6\pm0.9)\times10^{36}$~erg~s$^{-1}$.  
We use a 10 per cent uncertainty in the magnetic field strength for the three pulsars,
because of the uncertainty in the redshift correction and in the
measured energy of the cyclotron line.

\bsp	
\label{lastpage}
\end{document}